\documentclass[10pt,journal,final,twocolumn]{IEEEtran}

\makeatletter
\def\endthebibliography{%
	\def\@noitemerr{\@tablelatex@warning{Empty `thebibliography' environment}}%
	\endlist
}
\makeatother

\usepackage{epsfig,rotating,setspace,latexsym,amsmath,epsf,amssymb,amsfonts,bm,theorem,epstopdf} 

\usepackage{cite,caption,subcaption,enumitem,booktabs}
\usepackage{bbm}
\usepackage{algorithm}
\usepackage{algpseudocode}
\usepackage{color}

\usepackage[dvipsna9mes]{xcolor}
\usepackage{standalone}
\usepackage{tikz}
\usetikzlibrary{positioning}
\usetikzlibrary{shapes,arrows,chains}
\usetikzlibrary[calc]

\IEEEoverridecommandlockouts
\allowdisplaybreaks
\bstctlcite{IEEEexample:BSTcontrol}

\begin{document}
	
	\title{End-to-End Autoencoder Communications with Optimized Interference Suppression}
	
	\author{\IEEEauthorblockN{Kemal Davaslioglu, Tugba Erpek, and Yalin E. Sagduyu}
		\thanks{Distribution Statement A (Approved for Public Release, Distribution Unlimited). This research was supported under funding from Defense Advanced Research Projects Agency (DARPA). The views, opinions and/or findings expressed are those of the authors and should not be interpreted as representing the official views or policies of the Department of Defense or the U.S. Government.}			
	}	
	\maketitle

	\begin{abstract}
		An end-to-end communications system based on Orthogonal Frequency Division Multiplexing (OFDM) is modeled as an autoencoder (AE) for which the transmitter (coding and modulation) and receiver (demodulation and decoding) are represented as deep neural networks (DNNs) of the encoder and decoder, respectively. This AE communications approach is shown to outperform conventional communications in terms of bit error rate (BER) under practical scenarios regarding channel and interference effects as well as training data and embedded implementation constraints. A generative adversarial network (GAN) is trained to augment the training data when there is not enough training data available. Also, the performance is evaluated in terms of the DNN model quantization and the corresponding memory requirements for embedded implementation. Then, interference training and randomized smoothing are introduced to train the AE communications to operate under unknown and dynamic interference (jamming) effects on potentially multiple OFDM symbols. Relative to conventional communications, up to 36~dB interference suppression for a channel reuse of four can be achieved by the AE communications with interference training and randomized smoothing. AE communications is also extended to the multiple-input multiple-output (MIMO) case and its BER performance gain with and without interference effects is demonstrated compared to conventional MIMO communications.
	\end{abstract}
	\begin{IEEEkeywords}
		Autoencoder, deep learning, interference suppression, generative adversarial network, SISO, MIMO.
	\end{IEEEkeywords}
	
	\section{Introduction}
	Conventional communications systems typically separate the transmitter and receiver designs across individual communication blocks that are designed by analytical models or expert knowledge. Supported by recent advances of algorithmic and computational capabilities, deep learning provides an alternative to wireless communication tasks including the physical layer design by learning from and adapting to rich representations of spectrum data \cite{erpek1}. 
	
	\emph{Autoencoder} (AE) was first introduced in \cite{KramerAE} as a non-linear principal component analysis method. AE consists of two (deep) neural networks, an encoder that learns a latent representation (encoding) for a given set of data samples, typically for dimensionality reduction, and a decoder that aims to reconstruct the original data samples from this latent representation. An \emph{AE-based communications} system was introduced in \cite{oshea1} by \emph{jointly} training the transmitter and receiver functionalities. In particular, the transmitter and the receiver are designed as \emph{two deep neural networks} that represent the encoder and the decoder of the AE, respectively, while taking the channel effects into account instead of separately optimizing individual communication modules. Then, the goal is to optimize the transmitter and the receiver using an end-to-end approach by jointly training the encoder and the decoder of the AE to minimize the reconstruction loss that is translated to the bit error rate (BER) at the receiver. 
	
	In \cite{oshea1}, it was shown that the BER performance of AE communications either matches or exceeds that of conventional communications. In this formulation, the modulation scheme at the transmitter is optimized by the encoder training based on channel effects. In the meantime, the receiver is trained as a decoder to learn how to successfully demodulate the received signal. Designing capacity-approaching codes was considered in \cite{CapacityAppAE} with a novel loss function for the AE training. The AE communications over the air was shown with software-defined radios (SDRs) in \cite{Dorner1}. The single carrier approach for AE communications was extended in \cite{Felix1, aoudia1} to an Orthogonal Frequency Division Multiplexing (OFDM)-based system.

	In this paper, we consider an AE-based OFDM communications system, where the encoder represents the transmitter (namely, the coding and modulation operations) and the decoder represents the receiver (namely, the demodulation and decoding operations). While prior work has mostly focused on the adaptation of the AE communications to channel effects, our goal is to design the AE communications in a further practical scenario, where we consider \emph{training data limitations, implications of embedded hardware implementation, and suppression of external interference (such as jamming) effects} in addition to the channel effects.
	We measure the performance in terms of BER and error vector magnitude (EVM).  
	
	The AE communications is a data-driven approach and ultimately relies on the existence of representative training data to train the deep neural network structures of the AE. In particular, we show that the AE communications performance strongly depends on the number of available training samples. When limited training data is available (due to limited sensing time or sampling rate), we consider training a \emph{generative adversarial network (GAN)} \cite{goodfellow2014generative} to generate synthetic data samples and augment the training data for the AE communications system. Note that the more time is spent for data collection, the less time is left for data communications. We discuss how to select the appropriate number of synthetic data samples to optimize the BER performance. While a small number of synthetic data samples are not sufficient for the GAN to improve the BER performance through training data augmentation, too many synthetic data samples confuse the training via synthetic artifacts due to the data generation process.  
	
	Next, we consider the \emph{hardware constraints} on the AE communications. \emph{Embedded implementation} such as on field-programmable gate array (FPGA) or embedded Graphics Processing Unit (GPU) would be needed for edge devices (e.g., radio transceivers) and require computing operations implemented with finite precision. For that purpose, we \emph{quantize} the deep neural network (DNN) models and show that the quantization has only a very minor effect on the BER performance relative to the floating point implementation, while the memory requirement for the model size is significantly reduced by the quantization.
	
	Then, we introduce \emph{interference effects} and show how the AE communications can \emph{adapt to unknown and dynamic interference conditions} in addition to channel effects. For that purpose, we introduce a novel training method called \emph{interference training} to suppress the effects of the interfering signals. Also, we apply \emph{randomized smoothing} method that was introduced in \cite{Cohen} for computer vision applications to turn a classifier that classifies well under Gaussian noise into a new classifier that is certifiably robust to adversarial perturbations. To suppress interference effects, we train the AE communications using interference training and randomized smoothing. We first show that the AE communications outperforms the conventional communications when no interference such as a jamming signal is added yet. Then, we demonstrate the performance improvements (in terms of BER and EVM) that can be achieved with the proposed interference training and randomized smoothing methods under different interference conditions. We evaluate the cases where redundancy is added to the AE communications by using the channel multiple times for transmissions. When jamming signal is added as external interference, we evaluate the maximum jammer-to-signal-ratio (JSR) that can be tolerated while reaching a target BER under a given signal-to-noise-ratio (SNR). We show that \emph{up to $36$~dB interference suppression} for a channel reuse of four can be achieved with interference training and randomized smoothing. 
	
	We also extend AE communications for the single input single output (SISO) case to the multiple input multiple output (MIMO) case, where the size of the AE system increases with the number of transmit and receive antennas. For spatial multiplexing, we show that AE-based MIMO communications improves the BER significantly compared to conventional MIMO communications by accounting for channel effects across transmit and receive antennas. With interference training and randomized smoothing, we show that AE-based MIMO communications can effectively operate under interference effects achieving much lower BER compared to conventional MIMO communications.  
	
	\begin{figure}[t!]
		\centerline{\includegraphics[width=\linewidth]{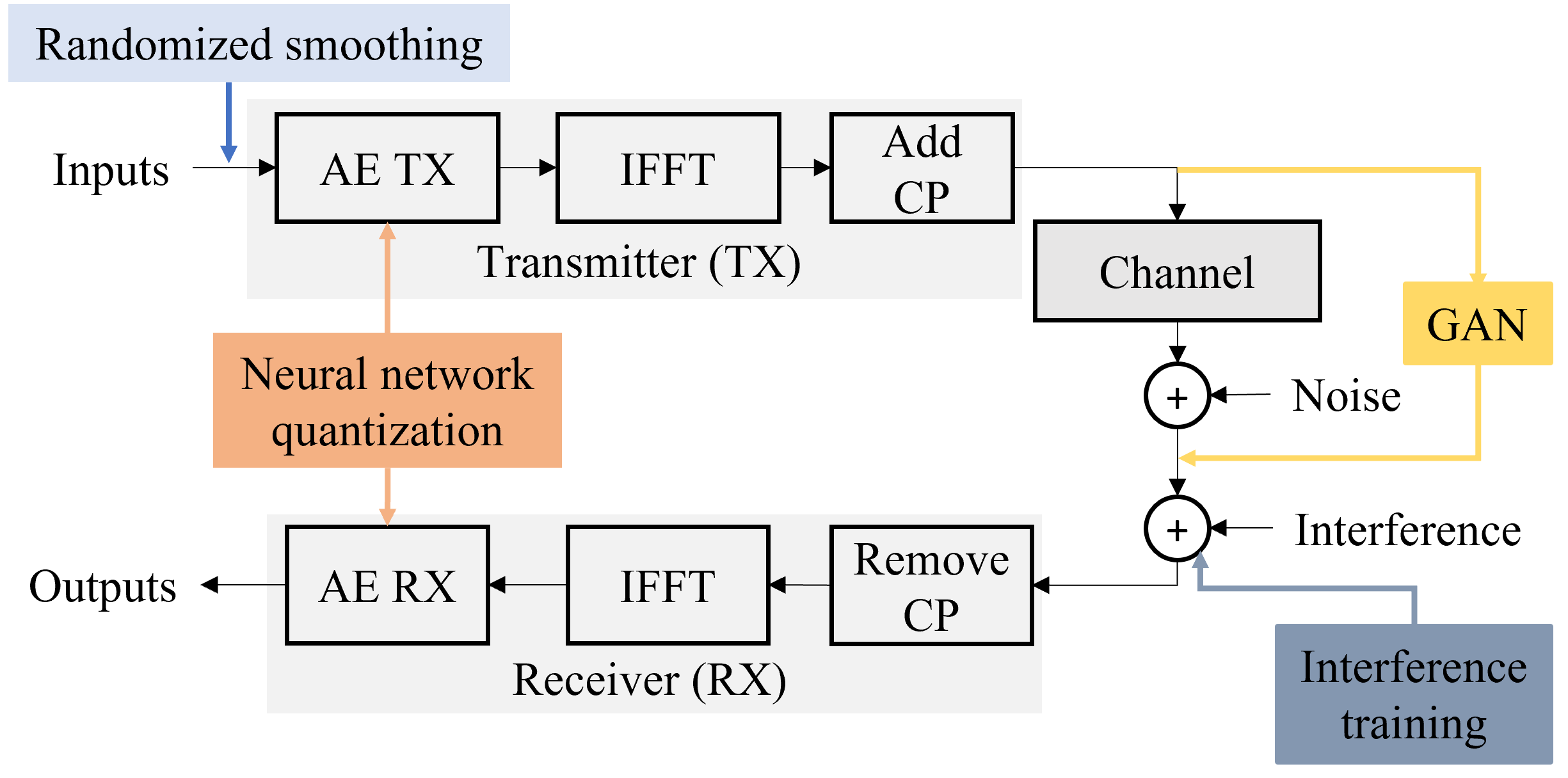}}
		\caption{Block diagram illustrating interfaces of AE communications with GAN-based data augmentation, quantization, randomized smoothing, and interference training components.}
		\label{fig:main_fig}
	\end{figure}
	
	Figure~\ref{fig:main_fig} shows the block diagram illustrating AE communications interfaces with GAN-based data augmentation, quantization, randomized smoothing, and interference training components. Our main contributions in this paper are summarized as follows:
	\begin{itemize}
		\item We develop an AE communications system that is designed to operate under unknown and dynamic interference (jamming) conditions. This approach uses a novel training procedure, called as interference training, and randomized smoothing. Our results demonstrate that the proposed solution can suppress the interference substantially more than the conventional communications. 
		\item We extend the effective capability of  AE communications for interference suppression from SISO to MIMO cases.
		\item We consider the case when there is not enough training data available to train the AE communications and develop a systematic approach that trains a GAN to augment the training data. We show that this approach improves the BER and EVM performance of the AE communications.
		\item We consider the constraints related to embedded hardware implementation that includes quantization and memory requirements, and show how the performance is sustained under the quantized model that reduces the corresponding memory requirements. 
	\end{itemize}
	
	The remainder of the paper is organized as follows. Section \ref{sec:RelatedWork} describes the related work. Section \ref{sec:SystemModel} presents the system model for the AE communications. Section \ref{sec:commperf} evaluates the AE communications performance under channel, quantization, and embedded implementation effects. Section \ref{sec:GAN} describes the use of the GAN for the AE training data augmentation. Section \ref{sec:int_train} introduces interference suppression to the AE communications via interference training and randomized smoothing. Section \ref{sec:MIMO} extends the AE communications with interference suppression to the MIMO case. Section \ref{sec:Conclusion} concludes the paper.
	
	\section{Related Work} \label{sec:RelatedWork}
	The use of the AE in wireless communications was introduced in \cite{oshea1} where it was shown that it is possible to learn the transmitter and receiver model for a given channel model. This approach was later extended to MIMO systems in \cite{Terpek17,SongGLOBECOM} and  massive MIMO systems in millimeter wave (mmWave) bands in \cite{hu2021two}. A channel-agnostic end-to-end communication system was developed in \cite{YeTCCN}, where the distributions of channel output are learned through a conditional GAN. Our work differs from this line of work in three major ways. First, our proposed solution trains the AE communications system with interference training and randomized smoothing for interference suppression along with adaptation to channel conditions, and considers limited training data and quantization effects. On the other hand, previous line of work focused on only the channel effects without specific consideration of external interference effects. Second, the typical approach as in \cite{oshea1} maps symbols to one-hot encoded vectors that become the input to the neural network. In our approach, we use bits as inputs and bits as outputs. Third, we use the binary cross-entropy (BCE) loss instead of mean squared error (MSE) loss (as used in \cite{oshea1}) to better represent the classification nature of the problem. 
	
	The idea of end-to-end learning of communications through DNN-based AEs was applied to OFDM systems in \cite{Felix1}. The performance of the AE communications was compared against that of a state-of-the-art OFDM baseline over frequency-selective fading channels in \cite{aoudia1}. In this setting, multiple channel uses are allowed for the transmission of a single message, thereby introducing redundancy to the transmissions in form of channel coding. 
	End-to-end learning was proposed in \cite{aoudia1} to design an OFDM communication system without cyclic prefix and pilot signals to decrease the overhead. Using a neural network, the transmitter learns a high-dimensional modulation scheme allowing to control the peak-to-average power ratio (PAPR) and adjacent channel leakage ratio (ACLR), as shown in \cite{goutay1}. The focus in \cite{goutay1} was on the mitigation of channel effects instead of achieving interference resistance. In addition, the joint optimization of the AE communications and beamforming through a reconfigurable intelligent surface (RIS) was considered in \cite{AE-RIS} by accounting for channel effects only.
	
	Interference channel was considered in \cite{ErpekIntf, Intf1} for single carrier communications systems where two encoder-decoder pairs that represent two concurrently operating transmitters-receiver pairs are jointly optimized in the presence of self-generated interference to minimize their symbol error rate (SER). In addition, interference management was considered at the receiver side by training DNNs in \cite{Intf2, Intf3, Intf4}. In this paper, we design the AE communications to suppress unknown and dynamic (random) interference effects that are externally added to the received signal in AE communications. 
	
	The previous work on the AE communications did not specifically consider model training while taking hardware limitations (e.g., at edge devices) on computing into account. In particular, training data limitations and quantization effects (as needed for embedded platforms such as in FPGA and embedded GPU) on AE communications were not considered in this previous work. Supervised machine learning techniques rely on the availability of training data. However, limited sensing time and low sampling rate may restrict the availability of training data. In that case, a GAN \cite{goodfellow2014generative} can be trained to generate synthetic data samples and augment training data. Training data augmentation has been considered for various wireless applications such as spectrum sensing \cite{Davaslioglu2}, jamming \cite{erpek2018deep,Shi2019generative}, and modulation classification \cite{clark2020training}. The GAN has been also used for channel modeling and channel estimation \cite{GAN1, GAN2, GAN3, GAN4}. In this paper, we apply the GAN to the AE communications, where synthetic data is generated by accounting for transmitter, receiver, and channel effects.

	\section{System Model} \label{sec:SystemModel}
\begin{figure}[h]
		\centerline{\includegraphics[width=\linewidth]{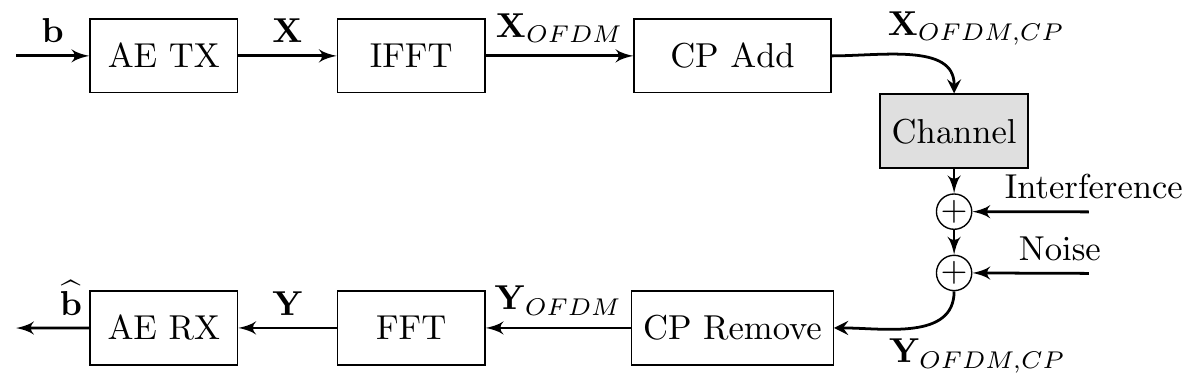}}
		\caption{AE communications system model.}
		\label{fig:system_model}
	\end{figure}

	The transmitter and the receiver are represented as two DNNs, one for the encoder and another one for the decoder, which jointly constitute an AE system as shown in Fig.~\ref{fig:system_model}. 
	
	The input to the transmitter is bits $\textbf{b}$ with size $n_{\text{FFT}}\times k$, where $n_{\text{FFT}}$ is the number of subcarriers and $k$ is the number of bits in a symbol. The output of the AE's encoder is the modulated symbols, denoted by \textbf{X} in Fig.~\ref{fig:system_model}. This architecture uses the \emph{number of channel uses} concept, similar to the one described in \cite{Felix1}, where redundancy is added to the transmitted symbols to improve performance in the form of joint modulation and coding. Multiple subcarriers as in OFDM communications are considered.   
	The channel use $n_{\text{ch}}$ refers to the number of times symbols are transmitted over the channel. Note that in the AE communications, different symbols can be transmitted in consecutive channel uses.  To achieve orthogonality, the modulated symbols go through the IFFT module. The input size to the IFFT module $\textbf{X}$ is $n_{\text{FFT}}\times k \times n_{\text{ch}}$, where $n_{\text{ch}}$ corresponds to the number of channel uses (redundancy). Cyclic Prefix (CP) is added to the output of the IFFT block $\textbf{X}_{\textit{OFDM}}$  to avoid inter-symbol interference (ISI) and CP added symbols are denoted as $\textbf{X}_{\textit{OFDM,CP}}$.

	The OFDM modulated symbols go through the channel effects before they are received at the receiver. Channel introduces frequency and phase shift to the input signal. Additive white Gaussian noise (AWGN) is added at the receiver. Also, unknown external interference signal (in particular jamming over different time samples) is added to introduce interference at a given JSR. The received symbols at the receiver are denoted by $\textbf{Y}_{\textit{OFDM,CP}}$. The receiver removes the CP and takes the FFT of the received signal. The resulting waveform $\textbf{Y}$ is the input to the AE's receiver module and bit estimates $\widehat{\textbf{b}}$ are output by the decoder. The encoder at the transmitter and the decoder at the receiver are jointly trained. Optimal transmit constellations are found at the end of the training based on the channel and interference effects with the goal of minimizing the BER. The AE is trained to minimize the difference between the input bits $\textbf{b}$ and the output bits $\widehat{\textbf{b}}$. This corresponds to a multi-label classification problem with the loss function given by 
	\begin{align} 
		\ell(\widehat{\textbf{b}},\textbf{b}) = & \frac{1}{N}  \sum_{n=0}^N \ell(\hat{b}_n,b_n)  = \frac{1}{N}  \sum_{n=0}^N \ell\left(h(b_n),b_n\right)  \label{eqn:bce}   \\
		& \hspace{-1.5em} = - \frac{1}{N}\sum_{n=0}^N \left( b_n \log (h(b_n)) + (1-b_n) \log (1 -h(b_n)) \right),
		\nonumber
	\end{align}
	where $b_n$ and $\hat{b}_n=h(b_n)$ represent the $n$th AE input and output bits, respectively, and $h(\textbf{b})$ is a function of the operations in the transmitter and receiver modules and intrinsically captures the impairments of channel and noise such that $\widehat{\textbf{b}}=h(\textbf{b})$. In our notation, we represent the total number of symbols transmitted over $n_{\text{ch}}$ channel uses by $N = (n_{\text{FFT}}+n_{\text{cp}})\times n_{\text{ch}}$, where $n_{\text{cp}}$ is the number of subcarriers for CP.
	The encoder and decoder DNNs of the AE architecture are shown in Table~\ref{table:autoencoder}.
	
	For numerical evaluation of the BER performance of the AE system, we assume $n_{\text{FFT}}=  12$ and $n_{\text{cp}} =3$ as an example. We vary the number of bits per symbol, while keeping $n_{\text{b}}$ as $2$ bits per symbol. The constellation points are determined by the DNNs depending on the channel and interference effects instead of having a fixed constellation such as in quadrature phase shift keying (QPSK) with $n_{\text{b}} = 2$.

	\begin{table}[tbh!]
		\caption{The AE architecture.}\label{table:autoencoder}
		\begin{center}
			\small
			\begin{tabular}{llc}
				\toprule
				Network & Layer & Properties \\ \hline
				Encoder & Linear  & $k \times n_{\text{FFT}} \times 4000$ \\
				& Activation  & ReLU \\
				& Dropout  & 25\% \\
				& Linear  & $4000 \times 2 \times n_{\text{FFT}} \times n_{\text{ch}}$ \\
				\midrule
				Decoder & Linear  & $2 \times n_{\text{FFT}} \times n_{\text{ch}} \times 4000$ \\
				& Activation  & ReLU \\
				& Dropout  & 25\% \\
				& Linear  & $4000 \times 4000$ \\
				& Activation  & ReLU \\
				& Dropout  & 25\% \\
				& Linear  & $4000 \times k \times n_{\text{FFT}}$ \\
				\bottomrule
			\end{tabular}
		\end{center}
	\end{table}
	
	\section{AE Communications Performance under Channel, Quantization, and Embedded Implementation Effects} \label{sec:commperf}
	
\begin{figure}[t!]
		\centerline{\includegraphics[width=\linewidth]{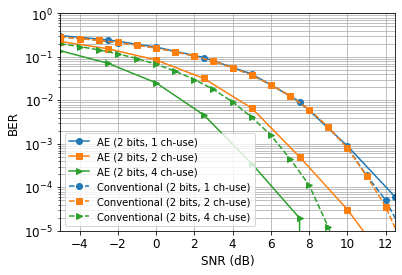}}
		\caption{SNR vs. BER for the AWGN channel.}
		\label{fig:snr_ber_awgn}
	\end{figure}

\begin{figure}[t!]
		\centerline{\includegraphics[width=\linewidth]{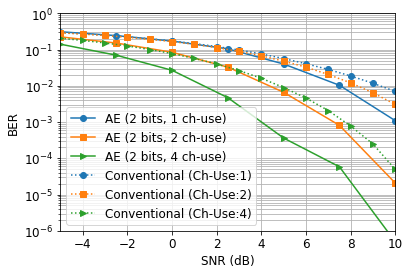}}
		\caption{SNR vs. BER for channel impairments in addition to the AWGN channel (phase offset is 10 degrees, frequency offset is 30 Hz).}
		\label{fig:snr_ber_ch_imp}
	\end{figure}

	The BER performance as a function of SNR for the AWGN channel is shown in Fig.~\ref{fig:snr_ber_awgn}, where we compare the AE communications with conventional communications. Note that in conventional communications, ``AE TX" in Fig~\ref{fig:system_model} is replaced with conventional modulation (depending on $n_{\text{b}}$) and repetition coding (transmitting the same symbol for $n_{\text{ch}}$ times), and ``AE RX" is replaced with the corresponding demodulation and decoding. The AE communications provides up to 3~dB improvement with increasing $n_{\text{ch}}$ at $10^{-3}$ BER.
	Fig.~\ref{fig:snr_ber_ch_imp} shows the BER as a function of the SNR when the channel impairments (phase offset is 10 degrees, frequency offset is 30 Hz) are introduced in addition to the AWGN channel. The performance gain of the AE communications relative to conventional communications is more apparent in the case with channel impairments. The reason is that as the added channel impairments distort the waveform, the performance of conventional communications degrades whereas the AE communications is trained to learn and compensate for these effects. The performance gain of the AE communications improves with increasing $n_{\text{ch}}$, since the AE jointly optimizes the channel use with modulation adding more flexibility for waveform design.

	Note that in Fig.~\ref{fig:snr_ber_awgn} for $n_{\text{ch}}=1$, we observe a slight degradation at SNR $>$ 12~dB when we compare the conventional and AE communications. However, when the effects of channel impairments are included, we observe the BER improvements around 2~dB even for $n_{\text{ch}}=1$ as shown in Fig.~\ref{fig:snr_ber_ch_imp}.

	
\begin{figure}[t!]
		\centerline{\includegraphics[width=\linewidth]{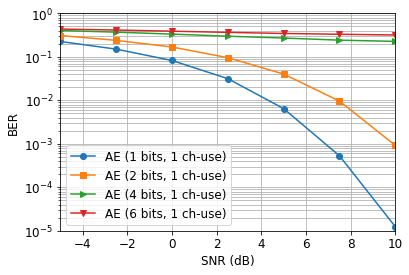}}
		\caption{SNR vs. BER for higher $n_{\text{b}}$ when $n_{\text{ch}}=1$.}
		\label{fig:snr_ber_higher_order_1ch}
	\end{figure}

\begin{figure}[t!]
		\centerline{\includegraphics[width=\linewidth]{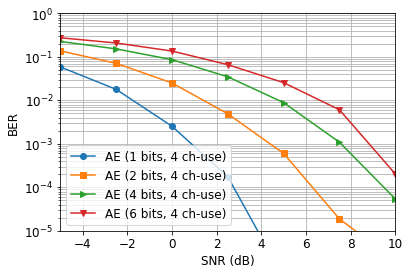}}
		\caption{SNR vs. BER for higher $n_{\text{b}}$ when $n_{\text{ch}}=4$.}
		\label{fig:snr_ber_higher_order_4ch}
	\end{figure}

	Next, we increase the number of bits for each symbol, $n_{\text{b}}$, and evaluate the performance of the AE communications. Fig.~\ref{fig:snr_ber_higher_order_1ch} shows the performance when $n_{\text{b}} = 1, 2, 4, 6$ for $n_{\text{ch}} =1$. Fig.~\ref{fig:snr_ber_higher_order_4ch} shows the  BER as a function of the SNR when $n_{\text{ch}}$ is increased to 4. A performance gain compared to conventional communications is observed for all cases with increasing $n_{\text{ch}}$. 
	
	\subsection{Constellation Comparison of the AE and Conventional Communications}
\begin{figure}[th!]
		\centerline{\includegraphics[width=\linewidth]{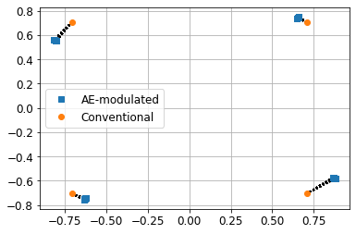}}
		\caption{Constellation plot of the AE and conventional communications at -5~dB SNR.}
		\label{fig:constellation_-5}
	\end{figure}

\begin{figure}[th!]
		\centerline{\includegraphics[width=\linewidth]{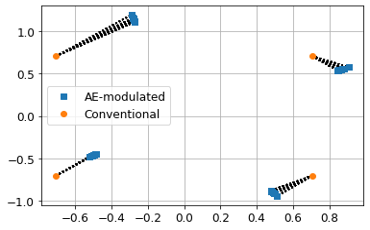}}
		\caption{Constellation plot of the AE and conventional communications at 30~dB SNR.}
		\label{fig:constellation_30}
	\end{figure}

	We evaluate the difference in constellations of the AE and conventional communications  at the transmitter (before IFFT and CP) when $n_{\text{ch}}=1$. The goal is to illustrate the difference of new constellations generated compared to conventional communications (QPSK in this case). The constellations are shown in Fig.~\ref{fig:constellation_-5} and Fig.~\ref{fig:constellation_30} for SNR is -5~dB and 30~dB, respectively. Note that the AE communications optimally selects the constellation to minimize the BER by adapting to channel conditions, whereas the conventional communication uses fixed constellation points.  
	
	\subsection{Error Vector Magnitude for Autoencoder Communications}
\begin{figure}[t!]
		\centerline{\includegraphics[width=\linewidth]{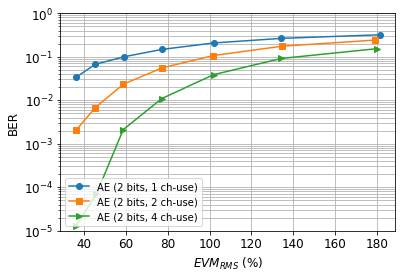}}
		\caption{EVM vs. BER for AE communications  with 2-bit per symbol when $n_{\text{ch}}=1, 2, 4$.}
		\label{fig:evm_2bit}
	\end{figure}

	Next, we measure the EVM (as the root mean square (RMS) average amplitude of the error vector normalized to ideal signal amplitude reference) for the AE communications. EVM is a measure of modulation quality and error performance in complex wireless environments and can determine the effects of impairments on signal reliability \cite{EVM}. For the AE communications, the error is measured as the difference from the AE encoder output at the transmitter to the AE decoder input at the receiver. The BER as a function of the EVM is shown in Fig.~\ref{fig:evm_2bit} for $n_{\text{b}} = 2$ and $n_{\text{ch}}=1, 2, 4$. Higher values of EVM indicate larger deviations between the ideal (reference) phasor and the received phasor and typically result in increased BER as shown in Fig.~\ref{fig:evm_2bit}. As we increase the channel use, coding gains help compensate for EVM effects and consequently the BER can be significantly reduced.

	\subsection{Quantization}
	In embedded implementation, the quantization operation reduces the resolution of bits and can create error propagation for the DNN due to rounding errors. Therefore, we investigate the effects of quantization on the BER performance. The DNNs are trained on a single NVIDIA GeForce RTX 2080 Ti GPU and quantized after training (post-training quantization). We use dynamic quantization that updates the bit resolution per neural network layer. While we may observe some performance degradation, quantization significantly reduces the model size and model weights need to be expressed in fixed point representation for FGPA and embedded GPU implementations. 
	
\begin{figure}[t!]
		\centerline{\includegraphics[width=0.495\textwidth]{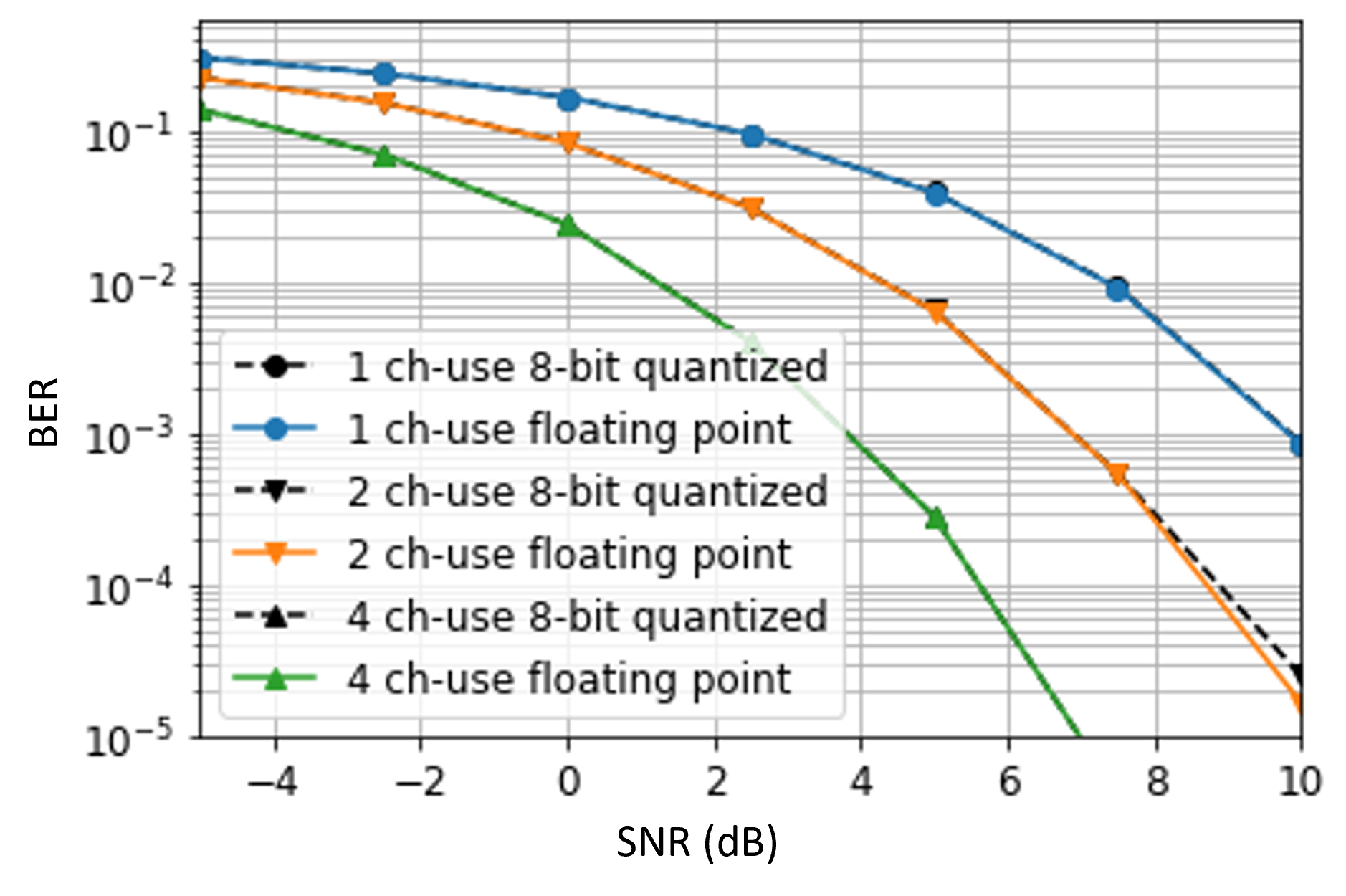}}
		\caption{BER performance comparison between the floating point model and 8-bit quantized model for $n_{\text{ch}}= 1, 2, 4$, respectively.}
		\label{fig:snr_ber_three_graphs_quantized}
	\end{figure}

	We study three performance measures: (i) model size, (ii) inference time, and (iii) error performance. Table~\ref{table:model_size} compares the model size of floating point model and 8-bit quantized models. We observe that the quantization reduces the model size by a factor of~4. Table~\ref{table:inference_time} presents the inferences time of floating-point model and 8-bit quantized model. The results are shown for per sample. A batch of 50 samples are evaluated and 10000 runs are averaged. We observe that the inference time of both models are very similar. The error performance of the floating point model and 8-bit quantized model are shown in Fig.~\ref{fig:snr_ber_three_graphs_quantized} 
	for $n_{\text{ch}} = 1, 2, 4$. Tables~\ref{table:quantization_chuse1}-\ref{table:quantization_chuse4} present the BER versus the SNR performance for different values of $n_{\text{ch}}$. We observe that quantization error is higher in low SNR cases (up to 0.0004) compared to the high SNR, but in all cases the effect of quantization is minor (less than $2.2 \%$).

	\begin{table}[h]
		\centering
		\caption{Model size comparison of floating point and 8-bit quantized models.}
		\label{table:model_size}
		\small
		\begin{tabular}{lll}
			\toprule
			Channel use & Floating point  & 8-bit quantized \\ $n_{\text{ch}}$
			& model & model\\
			\midrule
			1 & 2.88 Mbyte  & 0.73 Mbyte\\
			2 & 3.03 Mbyte  & 0.77 Mbyte\\
			4 & 3.34 Mbyte  & 0.85 Mbyte  \\\bottomrule
		\end{tabular}
	\end{table}

	\begin{table}[h]
		\centering
		\caption{Inference time comparison of floating point and 8-bit quantized models.}
		\label{table:inference_time}
		\small
		\begin{tabular}{lll}
			\toprule
			Channel use & Floating point  & 8-bit quantized \\
			$n_{\text{ch}}$ & model & model\\
			\midrule
			1 & 0.000109~sec  & 0.000114~sec\\
			2 & 0.000333~sec  & 0.000335~sec\\
			4 & 0.000354~sec  & 0.000358~sec  \\\bottomrule
		\end{tabular}
	\end{table}

	\begin{table}[t!]
		\centering
		\caption{BER performance comparison between the floating point model and 8-bit quantized model for $n_{\text{ch}}=1$.}
		\label{table:quantization_chuse1}
		\small
		\begin{tabular}{lll}
			\toprule
			SNR (dB) & Floating point  & 8-bit quantized \\
			& model & model\\
			\midrule
			-5.0 & 0.3033 & 0.3037 \\
			-2.5 & 0.2393 & 0.2394 \\
			0.0 & 0.1664 & 0.1665 \\
			2.5 & 0.0951 & 0.0952 \\
			5.0 & 0.0390 & 0.0392 \\
			7.5 & 0.0091 & 0.0093 \\
			10.0 & 0.0009 & 0.0009 \\
			\bottomrule
		\end{tabular}
	\end{table}
	
	\begin{table}[t!]
		\centering
		\caption{BER performance comparison between the floating point model and 8-bit quantized model for $n_{\text{ch}}=2$.}
		\label{table:quantization_chuse2}
		\small
		\begin{tabular}{lll}
			\toprule
			SNR (dB) & Floating point  & 8-bit quantized \\
			& model & model\\
			\midrule
			-5.0 & 0.2238 & 0.2242\\
			-2.5 & 0.1527 & 0.1529\\
			0.0 & 0.0831 & 0.0833 \\
			2.5 & 0.0308 & 0.0309 \\
			5.0 & 0.0064 & 0.0065 \\
			7.5 & 0.0005 & 0.0005 \\
			10.0 & 0.0000 & 0.0000 \\
			\bottomrule
		\end{tabular}
	\end{table}

	\begin{table}[t!]
		\centering
		\caption{BER performance comparison between the floating point model and 8-bit quantized model for $n_{\text{ch}}=4$.}
		\label{table:quantization_chuse4}
		\small
		\begin{tabular}{lll}
			\toprule
			SNR (dB) & Floating point  & 8-bit quantized \\
			& model & model\\
			\midrule
			-5.0 & 0.1376 & 0.1377 \\
			-2.5 & 0.0696 & 0.0697 \\
			0.0 & 0.0239 & 0.0240 \\
			2.5 & 0.0040 & 0.0041 \\		5.0 & 0.0003 & 0.0003 \\
			7.5 & 0.0000 & 0.0000 \\
			10.0 & 0.0000 & 0.0000 \\
			\bottomrule
		\end{tabular}
	\end{table}
	
	\subsection{Model size and latency measurements on FPGA for Low-Latency Embedded Applications}
	For embedded applications on edge radio devices, we study the model size and latency of the models in test time. The latency is measured via embedded implementation of the DNNs with FPGA. For that purpose, Xilinx Zynq Ultrascale MPSoC FPGA (XCZU7EV) is used. The software models (the encoder for the transmitter and the decoder for the receiver) are converted to the FPGA readable format and the resulting binary codes for the DNNs are ported to the FPGA following the approach in \cite{FPGA}. Table~\ref{table:fpga} shows the model size and number of parameters used in FPGA implementation. Per sample latency is measured as 8.69~microsecond at the transmitter and as 16.74~microsecond at the receiver, highlighting the effective use of the proposed approach for low-latency applications.
	
	\begin{table}[t!]
		\centering
		\caption{Model size and latency of the models used in FPGA implementation.}
		\label{table:fpga}
		\small
		\begin{tabular}{lll}
			\toprule
			Model & Model size  & Number of parameters \\
			\midrule
			AE-TX & 0.16~M  & 39,224\\
			AE-RX & 2.72~M  & 680,024\\
			\bottomrule
		\end{tabular}
	\end{table}

	\section{Generative Adversarial Network for AE Training Data Augmentation}\label{sec:GAN}
	
	The performance of the AE training depends on the number of available training samples. We study the case when there is limited training data available (e.g., when the time for channel sensing is short). We use a conditional data generation process with the GAN to learn the spectrum data distribution (received signals), generate synthetic data samples (signals), and augment the training data. 
	
	\begin{figure}[h]
		\centerline{\includegraphics[width=\linewidth]{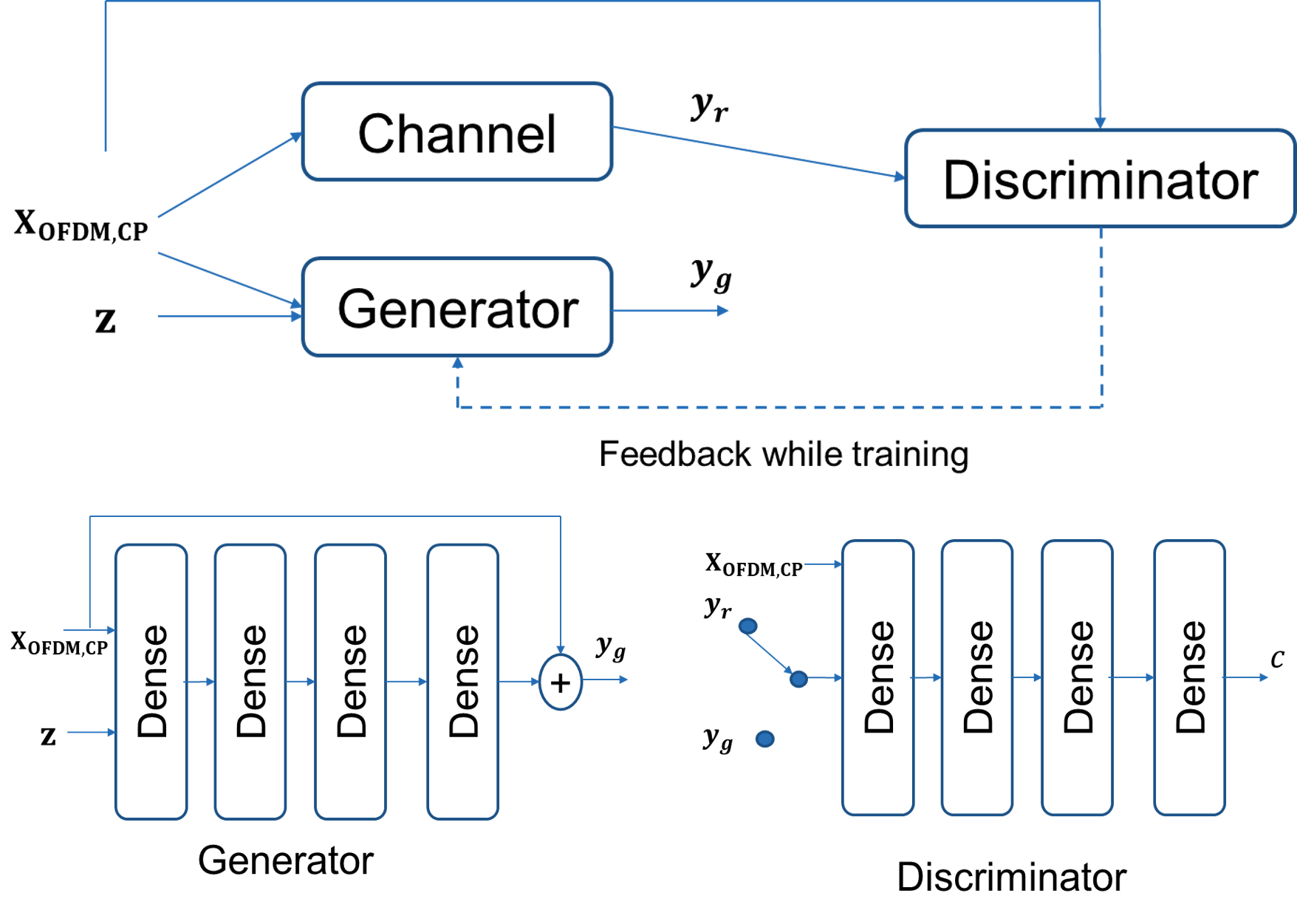}} 
		\caption{Conditional WGAN-GP system model.}
		\label{fig:gan_system_model}
	\end{figure}
	
	The system model of spectrum data learning using the GAN is shown in Fig.~\ref{fig:gan_system_model}, where the DNNs of the generator and the discriminator are jointly trained. The real data, represented by $\textbf{y}_r$ in Fig.~\ref{fig:gan_system_model}, includes the channel impairments with frequency and phase shifts. GAN learns the distribution of the real data over time and generates the synthetic data that is represented by $\textbf{y}_g$ in Fig.~\ref{fig:gan_system_model}. Typically, the GAN is trained using the minimax objective: 
	\begin{align} \label{eq:gan_loss}
		\begin{split}
			&\min_G \max_D \hspace{0.2in} \mathbb{E}_{\textbf{y}_r \sim \mathbb{P}_{\textbf{y}_r}} [\log(D(\textbf{y}_r))] \\ 
			& \hspace{0.68in} + \mathbb{E}_{\textbf{y}_g \sim \mathbb{P}_{\textbf{y}_g}} [\log(1 - D(\textbf{y}_g))],
		\end{split}
	\end{align}
	where $\textbf{y}_r$ represents the real data that follows the $\mathbb{P}_{\textbf{y}_r}$ distribution, $\textbf{y}_g=G(\textbf{z})$ is the synthetic data that uses  noise $\textbf{z}$ with a distribution $\textbf{z} \sim p(\textbf{z})$, $D(\cdot)$ is the output of the discriminator network, and $\mathbb{E}_\textbf{y}[\cdot]$ represents the expectation function over $\textbf{y}$. 
	When the discriminator reaches optimality before each generator network update, (\ref{eq:gan_loss}) reduces to minimizing the Jensen-Shannon divergence between real data distribution $\mathbb{P}_{\textbf{y}_r}$ and synthetic data distribution $\mathbb{P}_{\textbf{y}_g}$. However, the minimax objective in (\ref{eq:gan_loss}) typically leads to vanishing gradients as the discriminator saturates. To address this issue, we use the conditional Wasserstein GAN with gradient penalty loss (WGAN-GP) \cite{WGAN}. The objective function of conditional WGAN-GP is formulated as 
	\begin{align} \label{eq:wgan-gp}
		\begin{split}
			L = & \hspace{0.1cm} \mathbb{E}_{\textbf{y}_g \sim \mathbb{P}_{\textbf{y}_g}}[D(\textbf{y}_g| \textbf{x})] - \mathbb{E}_{\textbf{y}_r \sim \mathbb{P}_{\textbf{y}_r}} [D(\textbf{y}_r|\textbf{x})] \\
			& \hspace{.1in} + \lambda \cdot  
			\mathbb{E}_{\hat{\textbf{y}} \sim \mathbb{P}_{\hat{\textbf{y}}}} \left[ \left( || \nabla_{\hat{\textbf{y}}}   D(\hat{\textbf{y}}|\textbf{x}) ||_2 - 1 \right)^2 \right].
		\end{split}
	\end{align}
	The last term in (\ref{eq:wgan-gp}) is the gradient penalty, where $\hat{\textbf{y}}\sim\mathbb{P}_{\hat{\textbf{y}}}$ denotes the points that are uniformly sampled along the straight lines between the pair of points sampled from the generator distribution $\mathbb{P}_{\textbf{y}_g}$ and the real data distribution $\mathbb{P}_{\textbf{y}_r}$. The coefficient $\lambda$ represents the gradient penalty coefficient. The DNN architectures of the generator and discriminator are shown in Table~\ref{table:gan}.
	
	\begin{table}[tb!]
		\caption{Conditional WGAN-GP architecture.}\label{table:gan}
		\begin{center}
			\small
			\begin{tabular}{llc}
				\toprule
				& Layer name & Properties \\ \hline
				Generator & Linear  & $(20 + k \cdot n_{\text{FFT}}) \times 800$ \\
				& Activation  & ReLU \\
				& Linear  & $800 \times 800$ \\
				& Activation  & ReLU \\
				& Linear  & $800 \times 2(n_{\text{FFT}}+n_{\text{cp}}) n_{\text{ch}} $ \\
				
				\midrule
				Discriminator & 
				Linear  &  $4(n_{\text{FFT}}+n_{\text{cp}}) n_{\text{ch}}\times 64$ \\
				& Activation  & LeakyReLU, 0.2\\
				& Linear  & $64 \times 64$ \\
				& Dropout  & 40\% \\
				& Activation  & LeakyReLU, 0.2\\
				& Linear  & $64 \times 64$ \\
				& Dropout  & 40\% \\
				& Activation  & LeakyReLU, 0.2\\
				& Linear  & $64 \times 1$ \\
				\bottomrule
			\end{tabular}
		\end{center}
	\end{table}
	
	\subsection{Performance of Data Augmentation with GANs} \label{sec:results_gan}
	We evaluate two cases: 
	\begin{enumerate}[label=(\alph*)]
		\item \emph{Low-shot learning}: there is only limited amount of data to train the GAN and 100 real samples are used. 
		\item \emph{High-shot learning}: there is sufficient amount of data to train the GAN and 1000 real samples are used.
	\end{enumerate}	
	
	In both cases, first the GAN is trained and then the AE is trained. For the training of the AE communications, both the real samples and synthetic samples generated by the generator are used. It is important to determine the number of synthetic samples used for the AE training. Too many synthetic samples would lead to overfitting and too few of them would not help improve the performance. We evaluate the BER performance of AE communications with different number of synthetic samples. As the input to the generator of the GAN, a noise vector $\textbf{z}$ is used for the GAN training. The dimension of $\textbf{z}$ is selected as 50 (similar to \cite{Davaslioglu2}, where the GAN augmentation is used for spectrum sensing). Table~\ref{fig:ber_ratio} shows the BER performance for $n_{\text{ch}} =1 $ at 10~dB SNR when $N_{\text{real}}=100$ real samples are used and the number of synthetic samples, $N_{\text{synthetic}}$, is varied in $N_{\text{real}}$ increments. We observe that the number of synthetic samples added to the training set helps reduce the BER. For instance, when training the AE with $N_{\text{real}}=100$ real and $N_{\text{synthetic}} = 400$ synthetic samples (4x samples), the BER is reduced from 0.025 to 0.006. However, as we add more synthetic samples to the real data, the BER starts to increase since the AE model overfits to the GAN artifacts and deviates from the true (real) signal distribution.

	\begin{table}[ht!]
		\centering
		\caption{$N_{\text{synthetic}}$ vs. BER.}
		\label{fig:ber_ratio}
		\small
		\begin{tabular}{ll}
			\toprule
			$N_{\text{synthetic}}$ & BER \\
			\midrule
			No Augmentation & 0.025 \\
			2 $\times$ samples & 0.015 \\
			4 $\times$ samples & 0.006 \\
			6 $\times$ samples & 0.008 \\
			8 $\times$ samples & 0.012 \\\bottomrule
		\end{tabular}
	\end{table}
	
\begin{figure}[th!]
		\centerline{\includegraphics[width=0.495\textwidth]{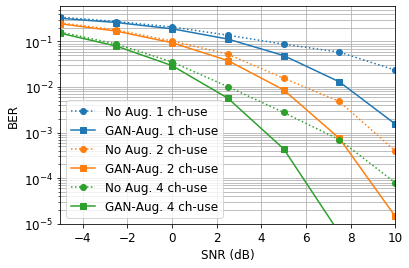}}
		\caption{BER performance with 100 real samples.}
		\label{fig:gan_100}
	\end{figure}

\begin{figure}[th!]
		\centerline{\includegraphics[width=0.495\textwidth]{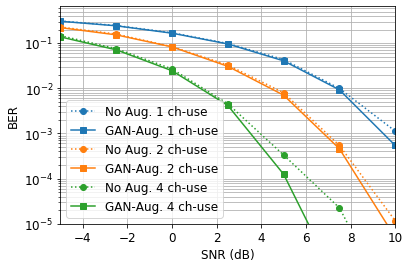}}
		\caption{BER performance with 1000 real samples.}
		\label{fig:gan_1000}
	\end{figure}

	To develop a systematic way of determining the number of synthetic data samples to add to training, the AE is trained with $N_{\text{real}}$ increments up to $10 \times N_{\text{real}}$ and the case with the best performance is selected. Fig.~\ref{fig:gan_100} shows the BER performance for the low-shot learning case where we compare the cases with and without augmentation for different values of $n_{\text{ch}}$. For example, data augmentation helps improve the BER performance (for $n_{\text{ch}} = 4$) at $10^{-4}$ from 10~dB to 6~dB providing a 4~dB improvement. Fig.~\ref{fig:gan_1000} depicts the data augmentation effects for the high sample case for different values of $n_{\text{ch}}$. We observe more modest gains when there are more training samples. For example, the BER only slightly improves for low SNRs, while we observe slightly higher gains at high SNRs (up to 2~dB improvement for $n_{\text{ch}} = 4$).

	\begin{figure*}[t!]
		\centering
		\begin{subfigure}[b]{0.32\textwidth}
			\centering
			\includegraphics[width=\textwidth]{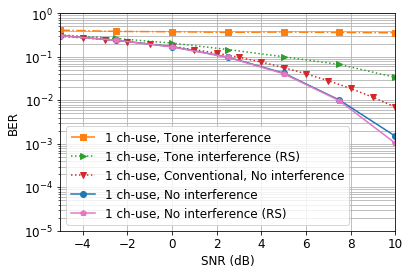}
			\caption{}
		\end{subfigure}
		\hfill
		\begin{subfigure}[b]{0.32\textwidth}
			\centering
			\includegraphics[width=\textwidth]{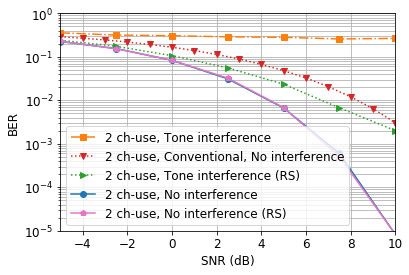}
			\caption{}
		\end{subfigure}
		\hfill
		\begin{subfigure}[b]{0.32\textwidth}
			\centering
			\includegraphics[width=\textwidth]{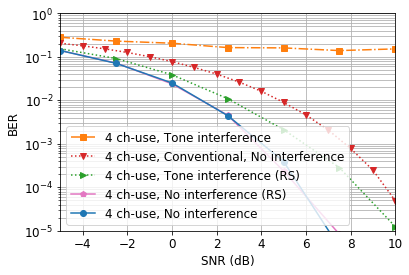}
			\caption{}
		\end{subfigure}
		\caption{SNR vs. BER when $n_{\text{ch}}$ is (a) $1$, (b) $2$, and (c) $4$, respectively.}
		\label{fig:snr_ber_three_graphs}
	\end{figure*}
	
	\section{Interference Suppression}\label{sec:int_train}
	
	Next, we describe the training for interference  suppression  that consists of the interference training and randomized smoothing steps. \emph{Interference training} anticipates for some level of interference during the training time such that the AE is trained with samples with and without interference. This approach reduces the disparity between the training and test data distributions under the presence of interference. Note that in test time, the AE does not know in advance the level of interference and its timing. First, we perform one regular training step that is a forward pass and a backpropagation for OFDM symbols without interference. This loss function can be expressed as
	\begin{align}
		\label{eq:reg_training_loss}
		\ell_{\text{AE}} = 
		\frac{1}{N}  \sum_{n=0}^N \ell\left(h(b_n),b_n\right).
	\end{align}		
	This step trains the DNN for interference-free clean signals (while accounting for channel effects). 
	
	In the next step, we add the interference effects and perform another forward pass and backpropagation. We repeat this process for several times. The loss function for the interference training is expressed as
	\begin{align}
		\label{eq:interference_training_loss}
		\ell_{\text{IT}} = \frac{1}{N_{it} N}\sum_{k=0}^{N_{it}}\sum_{n=0}^N \ell\left(g(b_n),b_n\right),
	\end{align}
	where $g(\textbf{b})=[g(b_1),\cdots,g(b_N)]$ is the function that includes the effects of both the channel impairments and  interference. Since there are different ways to create interference and interfere with (or jam) a signal of interest, we assume any OFDM symbol(s) can be randomly interfered with and this interference behavior is unknown to the communications design in advance. Later in this section, we investigate different cases where the AE communications is trained with $N_{\text{tr-jam}}$ symbol interference, but exposed to $N_{\text{te-jam}}$ symbol interference in test time, where $N_{\text{tr-jam}}$ is not necessarily equal to $N_{\text{te-jam}}$. 
	
	Second, we use \emph{randomized smoothing} (also referred to as Gaussian smoothing). This approach is initially proposed as a certified defense approach for computer vision applications to augment the training set with Gaussian noise or Laplacian noise and increase the robustness of the classifier against $\ell_2$ or $\ell_1$ attacks, respectively \cite{Cohen}. In this paper, we consider Gaussian noise $\textbf{n} \sim \mathcal{N}(\textbf{0},\sigma^2 \textbf{I})$ with a variance $\sigma^2$ to increase the robustness of the classifier to multiple directions. The loss function of randomized smoothing is given by
	\begin{align}\label{eq:rs}	
		\ell_{\text{RS}} = \frac{1}{N_{\text{rs}} N} \sum_{k=0}^{N_{\text{rs}}} \sum_{n=0}^{N} \ell (h(b_n + n_{k,n}),b_n)), 
	\end{align}
	where $n_{k,n}$ is the noise added to the input bit $n$ in the $k$th realization and the noise is added to all the bits in each realization. The noise vector $\textbf{n}_k = [n_{k,1},\cdots,n_{k,N}]$ is a zero-mean white Gaussian noise with variance $\sigma^2$. The parameters $N_{\text{rs}}$ and $\sigma^2$ help balance the accuracy and robustness tradeoff. Thus, we represent the final loss function that includes the contributions of the regular AE waveform training, interference training and randomized smoothing as
	\begin{align}\label{eq:total_loss}
		\ell_{\text{Total}} = \ell_{\text{AE}} + \ell_{\text{IT}} + \ell_{\text{RS}}.
	\end{align}
	
	\begin{algorithm}
		\caption{End-to-end AE training for interference suppression}\label{alg:cap}
		\begin{algorithmic}
			\State \textbf{Input}: To train without interference, $N$ and $h(b_n)$; for interference training, $N_{\text{it}}$ and $g(b_n)$; for randomized smoothing,  $N_{\text{rs}}$ and $\sigma^2$; and for AE optimization, the initial AE model $\boldsymbol{\theta}_0$, Adam hyperparameters $(\alpha, \beta_1, \beta_2)$
			\State \textbf{Initialize:} The first  moment vector $\textbf{m}_0 \leftarrow \textbf{0}$ and second moment vector $\textbf{v}_0 \leftarrow \textbf{0}$. 
			\For{epoch $t=1,2,\ldots, N_{\text{ep}}$}
			\For{$j = 1,2,\ldots, N_{\text{batches}}$}
			\State Generate random bits $\textbf{b}$ and calculate $h(\textbf{b})$.
			\State Compute AE loss without interference $\ell_{\text{AE}}$ with (\ref{eq:reg_training_loss}).
			\For{$k=1,\cdots,N_{\text{it}}$}
			\State Sample an interference vector.
			\State Calculate $g(\textbf{b})$.
			\State Compute interference training loss $\ell_{\text{IT}}$ with (\ref{eq:interference_training_loss}).
			\EndFor
			\For{$k=1,\cdots,N_{\text{rs}}$}
			\State Sample a noise vector $\textbf{n}_k \sim \mathcal{N}(0,\sigma^2 \textbf{I})$.
			\State Calculate $h(\textbf{b}+\textbf{n}_k)$.
			\State Compute 
			$\ell_{\text{RS}}$ using~(\ref{eq:rs}).
			\EndFor
			\EndFor
			\State Compute the gradient of (\ref{eq:total_loss}), i.e., $\boldsymbol{\rho}_t \leftarrow \nabla \ell_{\text{Total}}(\boldsymbol{\theta}_{t-1})$.
			\State Update biased first moment estimate \\ \hspace{1.5cm} $\textbf{m}_t \leftarrow \beta_1 \cdot \textbf{m}_{t-1} + (1-\beta_1) \cdot \boldsymbol{\rho}_t$.
			\State Update biased second moment estimate \\ \hspace{1.5cm} $\textbf{v}_t \leftarrow \beta_2 \cdot \textbf{v}_{t-1} + (1-\beta_2) \cdot \boldsymbol{\rho}_{t}^2$.
			\State Compute bias-corrected first moment estimate \\ \hspace{1.5cm} $\widehat{\textbf{m}}_t \leftarrow \textbf{m}_t /(1 - \beta_1^t)$.
			\State Compute bias-corrected second moment estimate \\ \hspace{1.5cm} $\widehat{\textbf{v}}_t \leftarrow \textbf{v}_t /(1-\beta_2^t)$.
			\State Update the model parameters with \\ \hspace{1.5cm} $\boldsymbol{\theta}_t \leftarrow \boldsymbol{\theta}_{t-1} - \alpha \cdot \widehat{\textbf{m}}_t /(\sqrt{\widehat{\textbf{v}}_t}+\epsilon)$.
			\EndFor
			\State \textbf{Output}: Final model $\boldsymbol{\theta}_k$.
		\end{algorithmic}
	\end{algorithm}
	
	The resulting end-to-end AE training for interference suppression is summarized in Algorithm~\ref{alg:cap}. The algorithm is trained for $N_{\text{ep}}$ episodes using $N_{\text{batches}}$ batches. For each batch, we generate $N$ random bits $\textbf{b}$. The bits are mapped to symbols using the encoder network, transmitted over the air where the effects of channel and noise, and hardware impairments are added. The received symbols are decoded using the decoder network. The $h(\textbf{b})$ includes all these effects. For randomized smoothing, we add a Gaussian noise $\textbf{n}$ to $\textbf{b}$ and calculate (\ref{eq:rs}). Finally, for interference training, we calculate $g(\textbf{b})$ that additionally includes the effects of interference. The loss in (\ref{eq:interference_training_loss}) is calculated. To update the AE model, Adam optimizer is used, where $\alpha$ is the learning rate, $\beta_1$ and $\beta_2$ denote the hyperparameters controlling the exponential decay rates of the moving averages for the first and second moments, respectively \cite{kingma2017adam}.
	
	We assume that interference can affect $N_{\text{te-jam}}$ symbols which are randomly selected (for each data sample) among all subcarriers used. For each selected symbol, we introduce a random phase shift that is uniformly randomly selected as $k \pi / 50$ where $k=0,\cdots,49$, while the magnitude of the energy of the interference signal is determined by the JSR.

	To evaluate the interference suppression capabilities of the AE communications, we introduce a tone jammer with a JSR of 15~dB in addition to channel effects and receiver noise. We consider three cases for AE communications:
	\begin{itemize}
		\item \textbf{Case~1}: No interference suppression. The AE communications is trained regularly as before. 
		\item \textbf{Case~2}: Interference suppression using randomized smoothing. 
		\item \textbf{Case~3}: Interference suppression using randomized smoothing and interference training. 
	\end{itemize}
	\noindent \textbf{Cases 1 and 2}: Randomized smoothing provides a robust way to suppress interference and closely maintains the BER performance when there is no interference effect. The BER as a function of the SNR is shown in Fig.~\ref{fig:snr_ber_three_graphs}(a)-(c), for $n_{\text{ch}} =1,2,4$, respectively. Interference from a time-domain jamming signal with a JSR of 15~dB is successfully suppressed such that the BER performance matches (or is better than) the conventional scheme (even without interference effects).

\begin{figure}[t!]
		\centerline{\includegraphics[width=\linewidth]{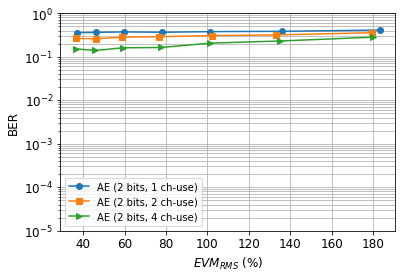}}
		\caption{EVM vs. BER when there is no interference suppression for AE communications evaluated under jamming conditions.}
		\label{fig:evm_ber_no_rs}
	\end{figure}

\begin{figure}[t!]
		\centerline{\includegraphics[width=\linewidth]{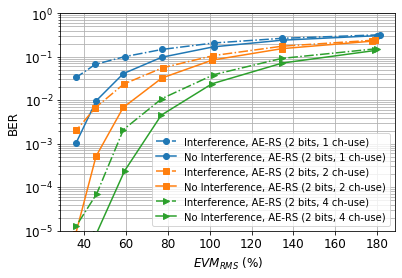}}
		\caption{EVM vs. BER when randomized smoothing is used for interference suppression.}
		\label{fig:evm_ber_int}
	\end{figure}

	Fig.~\ref{fig:evm_ber_no_rs} shows the BER as a function of the EVM for AE communications with no interference suppression, where we observe high BER even at the small EVM values. Fig.~\ref{fig:evm_ber_int} shows the BER as a function of the EVM for AE communications when it is trained using randomized smoothing. In this case, we observe that cases of $n_{\text{ch}} =2$ and $4$ maintain the BER of $10^{-2}$ at 50\% and 78\% EVM, respectively, under the 15~dB single-symbol interference. The results in Fig.~\ref{fig:evm_ber_int} clearly show the effectiveness of randomized smoothing as part of the AE communications to suppress the interference. 
	
\begin{figure}[ht!]
		\centerline{\includegraphics[width=\linewidth]{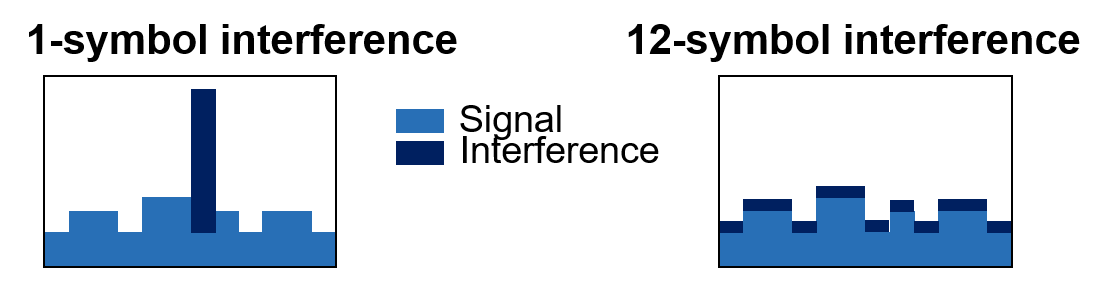}}
		\caption{Illustration of jamming power distributed to one symbol (left) and twelve symbols (right).}
		\label{fig:symbol_interference}
	\end{figure}

	\noindent \textbf{Case 3}: 
	In interference training, the AE is trained using a particular interference condition. For example, a model can be trained with 1-symbol interference. During test time, we evaluate the interference suppression performance of a model for a wide set of conditions that vary from 1-symbol interference to 12-symbol interference. For a fair comparison, we keep the total interference power the same when we consider $N$-symbol interference. This means that the test condition with 1-symbol interference allocates all the interference energy on a symbol per channel use, whereas 12-symbol interference means that the interference power is distributed across multiple symbols. Fig.~\ref{fig:symbol_interference} illustrates the difference between the two interference conditions. For simplicity, we consider equal power allocation across symbols.
	
	\begin{figure}[t!]
		\centering
		\begin{subfigure}[b]{\linewidth}
			\centering
			\includegraphics[width=\textwidth]{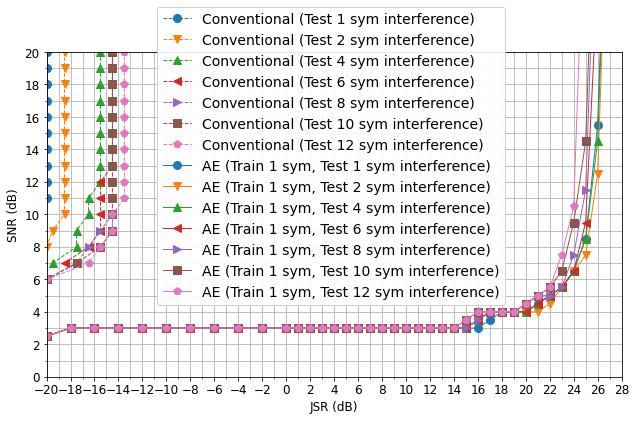}
			\caption{Trained on 1-symbol interference}
		\end{subfigure}
		\\
		\begin{subfigure}[b]{\linewidth}
			\centering
			\includegraphics[width=\textwidth]{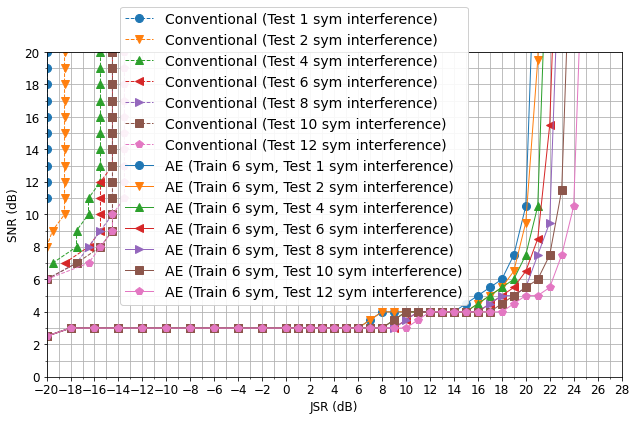}
			\caption{Trained on 6-symbol interference}
		\end{subfigure}
		\\
		\begin{subfigure}[b]{\linewidth}
			\centering
			\includegraphics[width=\textwidth]{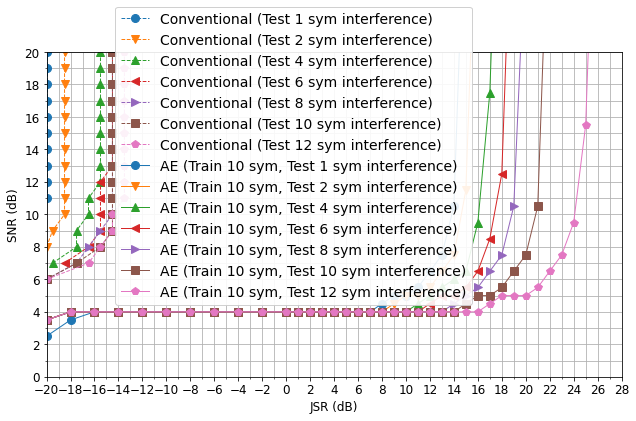}
			\caption{Trained on 10-symbol interference}
		\end{subfigure}
		\caption{Multi-symbol interference are evaluated for AE models trained with different interference conditions.}
		\label{fig:interference_train_multiple}
	\end{figure}

	To demonstrate the interference suppression capabilities, for each JSR value, we find the SNR value for which a BER less than or equal to the BER target of $10^{-2}$ is achieved leading to an achievable SNR-JSR plot. 
	Fig.~\ref{fig:interference_train_multiple}(a)-(c) shows the SNR-JSR plot for the AE models trained on 1, 6, and 10 symbols (i.e., $N_{\text{tr-jam}} = 1, 6, 10$) and compares their interference suppression capabilities with the conventional method under different test interference conditions (by varying $N_{\text{te-jam}}$). We observe that the AE models can suppress interference up to 24-26~dB. As the interference energy is distributed across more symbols, the AE model can tolerate higher JSR values. When compared to the conventional methods, the AE model with interference training and randomized smoothing provides more than 36~dB interference suppression.

	\section{Extension of AE Communications with Interference Suppression to MIMO} \label{sec:MIMO}
	In this section, we extend the AE-based SISO communications to AE-based MIMO communications for spatial multiplexing (independent streams transmitted from different antennas). Let $n_{\text{T}}$ and $n_{\text{R}}$ denote the number of transmit and receive antennas. 
	Dimensions of encoder and decoder inputs and outputs are given in Table~\ref{table:dimension}.
	
	\begin{table}[ht!]
		\centering
		\caption{Dimensions of encoder and decoder inputs and outputs for AE-based MIMO communications.}
		\label{table:dimension}
		\small
		\begin{tabular}{lllll}
			\toprule
			& Encoder & Encoder & Decoder & Decoder \\
			&  Input &  Output & Input &  Output \\
			\midrule
			Dimensions &	$n_{\text{b}} \times n_{\text{T}}$ & $2 n_{\text{T}}$ & $2 n_{\text{R}}$ & $n_{\text{b}} \times n_{\text{R}}$
			\\\bottomrule
		\end{tabular}
	\end{table}
	
	The FNN models for the encoder and decoder are:
	\begin{itemize}
		\item Encoder: Input-4000-4000-Output.
		\item Decoder: Input-4000-4000-4000-Output.
		\item In between each fully connected layer, we use a ReLU activation layer and a Dropout layer with probability 25\%.
	\end{itemize}
	
\begin{figure}[t!]
		\centerline{\includegraphics[width=\linewidth]{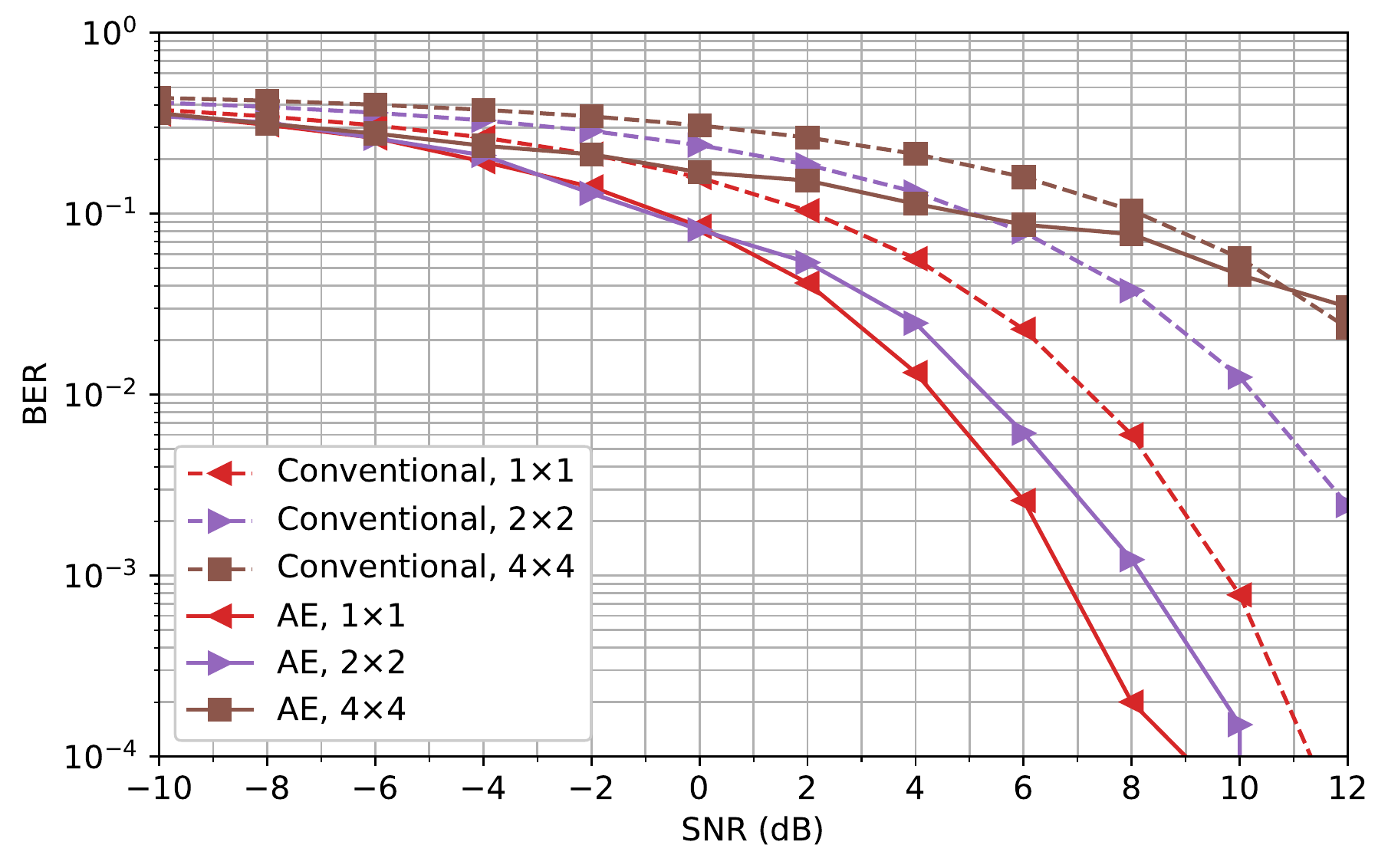}}
		\caption{The BER performance of AE-based vs. conventional communications (without interference) for the antenna configurations of $1 \times 1$, $2 \times 2$, and $4 \times 4$.}
		\label{fig:MIMO_BER}
	\end{figure}

	In terms of  channel impairments, we set phase offset as 10 deg and frequency offset as 30 Hz. Randomized smoothing is added to enrich the training data. The MIMO rate for spatial multiplexing increases with the number of antennas ($1\times 1$, $2 \times 2$, and $4 \times 4$). In the meantime, the AE-based MIMO communications reduces the BER compared to conventional MIMO schemes, as shown in Fig.~\ref{fig:MIMO_BER} when no interference is considered.
	
	\begin{table}[t!]
		\centering
		\caption{The MIMO BER performance under interference effects.}
		\label{table:MIMO}
		\small
		\begin{tabular}{llll}
			\toprule
			MIMO & BER for & BER for & BER for  \\
			Scheme &  $0$~dB SNR & $5$~dB SNR &  $10$~dB SNR \\ \midrule
			Conventional $2 \times 2$ &  0.2395 &  0.1048 & 0.0125 \\
			(No interference)&   &   & \\
			\midrule
			AE $2 \times 2$ &  0.1327 &  0.0489 & 0.0028 \\
			(interference)&   &   &  \\
			\midrule
			Conventional $4 \times 4$  &  0.3084 &  0.1867 & 0.0569 \\
			(No interference)&   &   &  \\
			\midrule
			AE $4 \times 4$ &  0.1929 &  0.1244 & 0.0405  \\
			(Interference)&  &  & 
			\\\bottomrule
		\end{tabular}
	\end{table}
	
	Next, we introduce interference effects in addition to channel effects and evaluate how AE-based MIMO communications can suppress the interference. For that purpose, MIMO AE communications is trained using randomized smoothing and interference training. Interference conditions are set as 1-symbol (randomly selected) interference at $0$~dB JSR, and 2 bits/symbol ($n_{\text{b}} = 2$) and 1 channel use ($n_{\text{ch}} =1$) are employed.
	
	Table~\ref{table:MIMO} shows that the AE communications improves the BER for MIMO compared to conventional communications when interference is introduced. The BER improvement by the AE-based MIMO communications is consistent across different SNRs and antenna configurations. In particular, compared to conventional communications (even when no interference is added), the AE communications (subject to interference) reduces the BER by $59\%$ on average for the $2 \times 2$ MIMO case and by $33\%$ on average for the $4 \times 4$ MIMO case.
	
	\section{Conclusion} \label{sec:Conclusion}
	We considered an end-to-end communications system (based on OFDM) that is modeled as an AE for which the transmitter and receiver  are represented by DNNs modeled as the encoder and decoder, respectively. First, we showed that the AE communications improves the BER performance of the conventional communications for the SISO case. In particular, we considered practical scenarios, where (i) the number of training data samples is limited, (ii) there are embedded implementation limitations, and (iii) communication is subject to both channel and interference effects. We trained a GAN to augment the training data for AE communications and show the improvement in the BER performance. We observed that the effect of the DNN model quantization is minor for the BER performance whereas the quantization reduces the memory requirement for the embedded implementation.  Then, we designed the AE communications to operate in the presence of an unknown and dynamic interference (jamming) conditions. Our solution  based on interference training and randomized smoothing achieves up to 36~dB interference suppression relative to conventional communications. We also extended the AE communications to the MIMO case. For spatial multiplexing, we showed that the AE communications improves the BER performance of the conventional communications for different antenna configurations. Then, we introduced interference effects and showed that interference training and randomized smoothing are effective in suppressing interference also when AE is used for MIMO communications. 
	
	\bibliographystyle{IEEEtran}
	\bibliography{IEEEabrv,references}
	
\end{document}